# Long Secondary Periods in Red Giants: AAVSO Observations and the Eclipse Hypothesis

*John R. Percy and Melanie Szpigiel*

Department of Astronomy & Astrophysics, and Dunlap Institute for Astronomy and Astrophysics, University of Toronto, 50 St. George Street, Toronto ON M5S 3H4; john.percy@utoronto.ca

**Subject Keywords**

AAVSO International Database; Photometry, CCD; Photometry, visual; Long-period variables; Semi-regular variables; Period analysis; Amplitude analysis; Stars: individual (SS Cep, AF Cyg, U Del, EU Del, TX Dra, UW Her, OP Her, Y Lyn, W Ori, W Tri, ST UMa)

**Abstract**

At least a third of red giants show a long secondary period (LSP), 5 to 10 times longer than the pulsation period. There is strong evidence that the LSP is caused by eclipses of the red giant by a dust-enshrouded low-mass companion. We have used long-term AAVSO observations of 11 stars to study two aspects of the eclipse hypothesis: the relation between the LSP phase (eclipse) curve and the geometry of the eclipse, and the long-term (decades) changes in the LSP phenomenon in each star. The stars with the largest LSP amplitudes show evidence of a dust tail on the companion, but most of the 11 stars show only a small-amplitude sinusoidal phase curve. The LSP amplitudes of all the stars vary slowly by up to a factor of 8, suggesting that the amount of obscuring dust varies by that amount, but there is no strong evidence that the geometry of the system changes over many decades.

## 1. Introduction

Red giant stars are unstable to low-order pulsation modes, generally the fundamental and/or first overtone mode. About a third of such stars also have a "long secondary period" (LSP), 5 to 10 times the pulsation period, depending on whether the pulsation is in the fundamental or first overtone mode. The cause of these LSPs was unknown for almost a century. Recently, Soszynski et al. (2021) have presented strong evidence that they are due to eclipses of the red giant by dust-enshrouded companions which were originally planets, but which subsequently accreted gas and dust from the star and became brown dwarf or low-mass stars. Figure 1 shows an artist's conception of an LSP system.

A simple model of the system would then include the red giant (probably pulsating), the low-mass companion with a halo of dust and gas, perhaps trailing a dust tail which merges with a uniform (or not) ring of gas and dust around the red giant. But, as Figure 1 suggests, the situation can be more complicated. The outer layers of the red giant are not uniform and symmetric, but consist of large, random convection cells. The pulsation amplitude of the red giant varies by up to a factor of ten, so the gas and dust flows are patchy. The system may be seen at any angle, not necessarily edge-on; this would affect the interpretation.

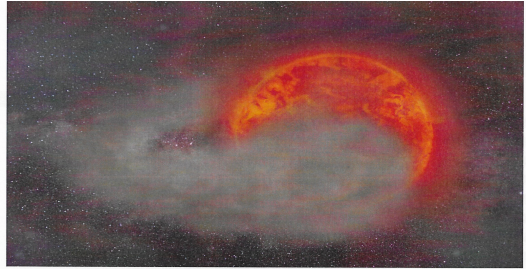

*Figure 1. Artist's conception of the LSP process. Source: Matylda Soszynska*

Despite the success of the Soszynski et al. mechanism for explaining the LSP phenomenon, there are still some puzzling aspects, and those have been the topic of some previous papers by our group. The present paper deals with two aspects of the LSP process: (1) the relationship between the LSP (eclipse) phase curve and the geometry of the eclipse process, and (2) changes (if any) in the LSP process in the star over a period of decades. We can study such changes because of the existence of the American Association of Variable Star Observers (AAVSO) International Database (AID) of visual and photoelectric photometry, which extends back in time for many decades. We have previously used some of this data in several studies of LSPs in red giants (e.g. Percy and Diebert (2016)), but we realized that there was even more information which could be derived.

*1.1 Eclipse phase curve and the eclipse geometry*

Let us consider how the LSP system, as seen in Figure 1, produces the observed light and phase curves. If the LSP system is seen flat-on, there will be no eclipses. If the system is seen at an angle – 45 degrees, for instance – the dusty companion may eclipse the top or bottom of the face of the red giant, producing a relatively short "partial" eclipse with a flat maximum. If the system is seen edge-on, there are several possibilites. If the companion is small and opaque, the minima and maxima will both be flat. The depth of the eclipse will depend on the ratio of the areas of the two objects. The length of the eclipse will depend on the relative sizes of the red giant and its companion. If the companion and its halo are large and relatively transparent, the eclipse will be slow and symmetric; the egress will be a reflection of the ingress. If the companion has a dust tail behind it, the ingress will be relatively shorter and the egress with be longer because of absorption by the tail, as seen in Figure 2. Figure 1 shows a tail which is extended into a disc or torus, which may completely encircle the red giant. In that case, the disc of the red giant may never be completely unobscured.

Figure 1 in Soszynski et al. (2021) and the Introduction to that paper are instructive. The figure shows eclipse (LSP) phase curves for a selection of stars in the Galactic bulge and LMC. The smallest-amplitude stars have short, rounded eclipses and flat maxima. The largest-amplitude stars have a more rapid ingress, a slower egress, and pointed minima. Soszynski and Udalski (2014) had demonstrated that such a phase curve could be produced by a dusty cloud with a comet-like tail. One caveat: the stars in this figure are chosen to have small pulsation amplitudes, to highlight the LSP variations; this may introduce some bias in the shape of the eclipse curves shown, such as preferentially including lower-luminosity stars.

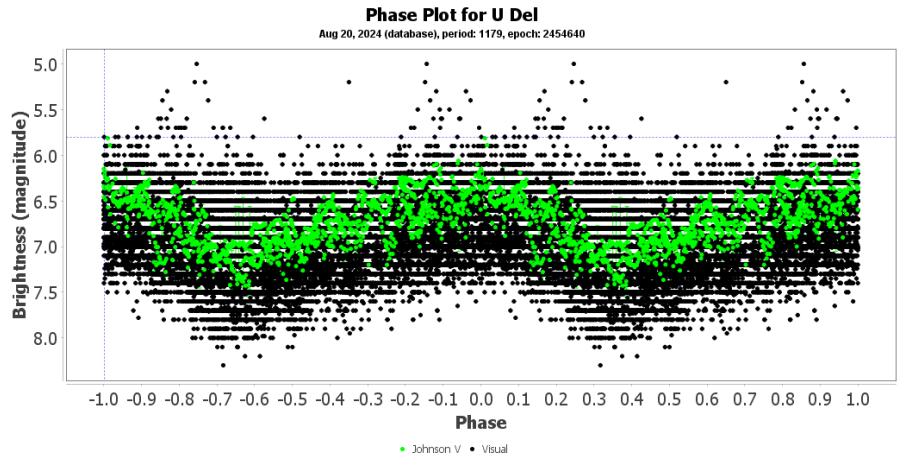

*Figure 2. The phase (eclipse) curve for U Del. The black points are visual, the green points are Johnson V. The LSP is 1178 days. Source: AAVSO.*

*1.2 Long-term changes in the LSP phenomenon*

Visual data in the AID goes back almost a century for some red giants; photoelectric Johnson V photometry extends about half as far back. To study long-term changes in the LSP process, we require stars for which the data are dense and sustained. We drew most of the stars from a recent compilation by Percy and Zhitkova (2023), which analyzed data from the AAVSO Binocular Observing Program. One problem is that the pulsational variations act as "noise" in studying the LSP variations, adding to the usual observational error noise. Table 1 lists the 11 stars that we considered suitable for our study. The last column lists the ratio of the LSP amplitude LSPA to the pulsation amplitude PA. This is a measure of the "signal-to-noise" for studying the LSP variation.

Why does all this matter? Because red giants are common; the AAVSO monitors hundreds of them. The sun will become a red giant; will it exhibit the LSP process? There is much interesting physics in the processes by which the red giant produces dust, and its dust-enshrouded companion. And LSPs are a stellar mystery which AAVSO data have helped to solve.

**2. Data and Analysis**

Visual and Johnson V observations of the 11 stars in Table 1 were downloaded from the AID, and analyzed using careful light-curve analysis, and time-series analysis with the AAVSO VSTAR software package (Benn 2013). The LSPs and pulsation periods, and their amplitudes were taken from Percy and Zhitkova (2023) and checked, if necessary, with the Fourier routine in VSTAR. Note that the amplitudes given by VSTAR, and quoted in this paper, are actually semi-amplitudes, not max-to-min ranges.

*Table 1. Red giants with LSPs and sustained observations in the AID. The table gives the starname, LSP period and mean, maximum, and minimum amplitude, pulsation period and amplitude, ratio of LSP to P, and ratio of LSP amplitude to pulsation amplitude.*

| Star | LSP(d) | LSPA | LSPAmax | LSPAmin | PP(d) | PA | LSP/P | LSPA/PA |
|---|---|---|---|---|---|---|---|---|
| SS Cep | 947 | 0.11 | 0.21 | 0.07 | 170 | 0.17 | 5.57 | 0.65 |
| AF Cyg | 913 | 0.08 | 0.17 | 0.02 | 93 | 0.18 | 9.82 | 0.44 |
| U Del | 1179 | 0.21 | 0.26 | 0.14 | 120 | 0.08 | 9.83 | 2.63 |
| EU Del | 630 | 0.05 | 0.10 | 0.03 | 63 | 0.10 | 10.00 | 0.50 |
| TX Dra | 700 | 0.14 | 0.19 | 0.10 | 76 | 0.12 | 9.21 | 1.17 |
| UW Her | 986 | 0.07 | 0.11 | 0.06 | 182 | 0.17 | 5.42 | 0.41 |
| OP Her | 699 | 0.10 | 0.14 | 0.03 | 74 | 0.06 | 9.45 | 1.67 |
| Y Lyn | 1245 | 0.35 | 0.44 | 0.15 | 135 | 0.09 | 9.22 | 3.89 |
| W Ori | 2358 | 0.20 | 0.23 | 0.18 | 432 | 0.13 | 5.46 | 1.60 |
| W Tri | 768 | 0.07 | 0.10 | 0.06 | 109 | 0.04 | 7.05 | 1.75 |
| ST UMa | 615 | 0.06 | 0.14 | 0.04 | 90 | 0.13 | 6.83 | 0.46 |

The visual observations were analyzed preferentially, since they were more numerous and extend back the furthest. Note that the visual and V passbands are slightly different, so the V observations are offset from the visual ones, as you can see in the light and phase curves. Eclipse phase diagrams were constructed with VSTAR (e.g. Figure 2). For some of the stars, the shape of the LSP phase curve was apparent from visual inspection. In addition, the phase curves could be fitted with low-order (typically 4) polynomials, using the polynomial function in VSTAR. The shape of the phase curve could be then described, very simply, by the phase difference between maximum and minimum – or the fraction of the LSP cycle in which the star is decreasing in brightness. If the phase curve was symmetrical, this parameter would be 0.50. If the ingress was faster that the egress, it would be less than 0.50. The depth of the eclipse could be measured from the difference between maximum and minimum of the fitted polynomial. For stars with dense, sustained data, changes in these parameters with time could be determined by dividing the data into time segments and analyzing each segment separately. Changes in the depth of the eclipse could also be determined using the wavelet routine within VSTAR. A helpful strategy for detecting cycle-to-cycle changes in some of the LSP light curves was to average the measurements in bins equal to the pulsation period. It cuts down, partially, on the pulsation "noise".

## 3. Results

The 11 stars in our sample are listed in Table 1, along with some results. Notes on individual stars follow:

- SS Cep: LSPA is only 0.11; PA is 0.17. The former varies by a factor of three. Early in the dataset, the phase curve seems to show a slower decline to minimum and a faster return to maximum; later in the dataset, the phase curve is more sinusoidal.

- AF Cyg: LSPA is less than half PA, so there is significant pulsation "noise" in the LSP analysis. The LSPA variation – 0.02 to 0.17 – is proportionally the largest in our sample. The phase curve is small-amplitude, sinusoidal, and essentially constant over the period of the dataset.
- U Del: LSPA is relatively large, and PA is relatively small, so the LSP is easier to analyze and interpret; it is one of our two best cases, along with Y Lyn. The phase curve (Figure 2) has a distinctive shape, with a more rapid decline to minimum and a slower return to maximum; this would be consistent with a tail following the companion as it eclipses the red giant. Also, the star spends about half the LSP at or near maximum, which suggests that the star may be relatively unobscured in this interval. Early in the dataset, the phase curve was slightly more sinusoidal; this may be because the LSP amplitude was lower at that time (Figure 3).
- EU Del: LSPA is small; PA is twice as large. This star has extensive photoelectric coverage, and is aprototype for SARVs: small-amplitude red variables. It is much easier to analyze the pulsation than the LSP. LSPA varies by a factor of three. The polynomial fit suggests that the star has a slow decline to minimum and a more rapid rise to maximum, but the effect is very small and is affected by the strong pulsational noise.
- TX Dra: LSPA and LSPA/PA are moderate. LSPA varies by a factor of two. The shape of the phase curve remains uniformly sinusoidal over the period of the dataset.
- UW Her: LSPA is moderate, but PA is relatively large (0.17) so the pulsation "noise" level is high. LSPA varies from 0.06 to 0.11, but the phase curve remains uniformly sinusoidal over the period of the dataset.
- OP Her: LSPA and LSPA/PA are moderate. There is a large range in LSPA – from 0.03 to 0.14 – but the phase curve is uniformly sinusoidal over the course of the data.
- Y Lyn: LSPA is large, and much larger than PA so, not surprisingly, this was the easiest star to analyze and interpret. This star's phase curve (Figure 4) has the same distinctive properties of that of U Del (see above) and can be interpreted in the same way -- in terms of a trailing dust tail.
- W Ori: LSPA is small, but PA is smaller. Because of the longer length of LSP and P, it is easy to see the effect of the pulsation "noise" in the phase curve.
- W Tri: LSPA is small, but PA is smaller. LSPA varies from 0.06 to 0.10; the phase curve remains sinusoidal over the course of the data (Figure 5), typical of many of the stars in our sample.
- ST UMa: LSPA and LSPA/PA are both small, so the data are difficult to analyze and interpret. LSPA varies from 0.04 to 0.14; the phase curve remains sinusoidal over the period of the dataset.

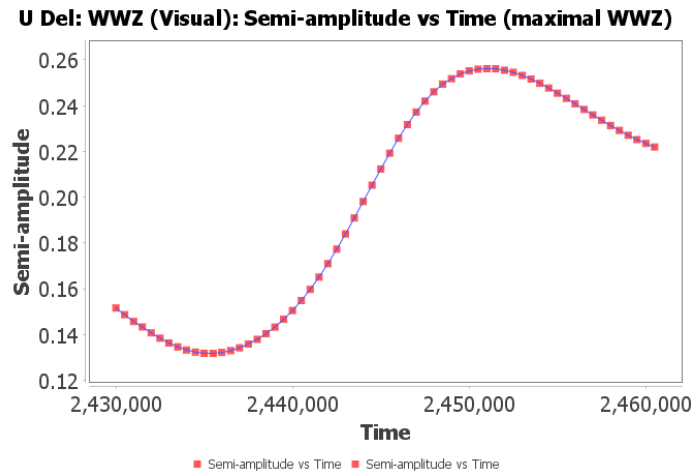

*Figure 3. The semi-amplitude of the U Del LSP – which is a measure of the depth of the eclipse – versus time. Source: AAVSO.*

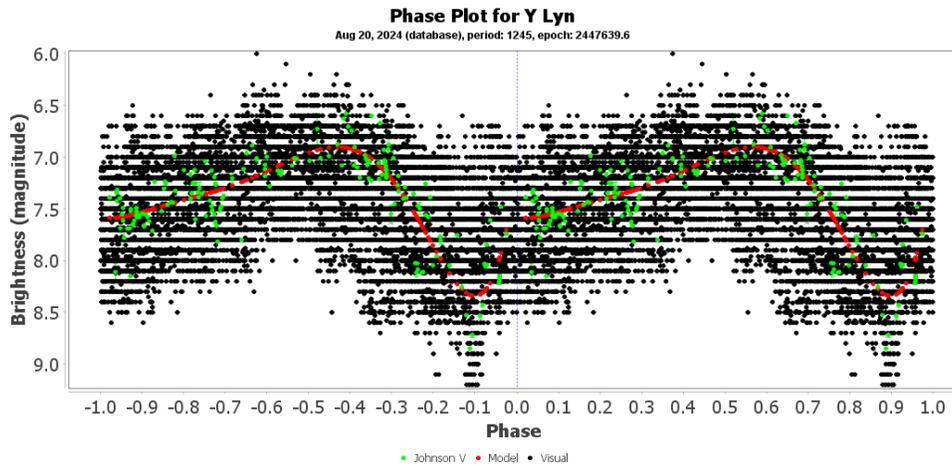

*Figure 4. The phase (eclipse) curve for the LSP in Y Lyn. The black points are visual, the green points are Johnson V, and the red curve is a fitted fourth-degree polynomial. The LSP is 1245 days. Source: AAVSO.*

### 4. Discussion and Conclusions

Mass loss in Mira stars is usually associated with their pulsation. It is somewhat puzzling that, in stars with small pulsation amplitudes -- U Del and Y Lyn for example – the star can drive off sufficient dust and gas to increase the companion mass, and produce its envelope of dust, and its tail.

The pulsational "noise", on top of the observational noise make the analysis in this project difficult. For many of the stars in Table 1, the signal-to-noise for the LSP analysis is less than one. It might be worthwhile to repeat this analysis using only V data on stars with dense photoelectric/CCD coverage, even though that data extends back by only 3-4 decades.

It is somewhat puzzling that, in stars with small pulsation amplitudes -- U Del and Y Lyn for example – the star can drive off sufficient dust and gas to increase the companion mass, and produce its envelope of dust, and its tail.

Nevertheless, the relatively small scatter in the phase curves for U Del and Y Lyn – especially the V data -- supports the conclusion that the phase curse, and the eclipse geometry do not change significantly over a few decades.

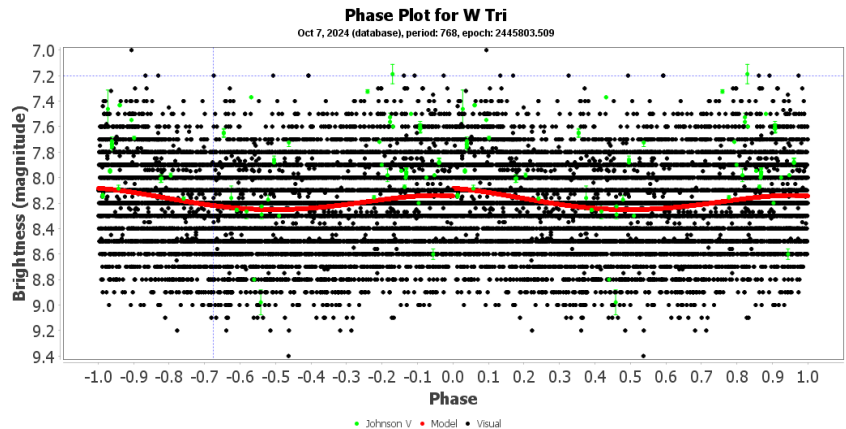

*Figure 5. The phase (eclipse) curve for W Tri. This, with its small amplitude and sinusoidal shape, is typical of over half the stars in our sample. The red curve is a polynomial fit. Source: AAVSO.*

The amount of obscuring dust, as indicated by the LSP amplitude, can vary by a factor of up to eight, and does so on a time scale of decades. The eclipse geometry, as reflected by the uniformity of the light curve, and the relatively small scatter in the phase (eclipse) curve, appears not to change significantly over the time interval of our data, despite the dust variation. At least two of our larger-LSPA stars show an asymmetrical phase curve which can be interpreted as due to a dust tail following the companion. For most of our stars, however, the LSP amplitude is small, and the phase curve is sinusoidal. None show "classical" eclipse curves with flat maxima.

**Acknowledgements**

We acknowledge with thanks the variable star observations from the AAVSO International Database contributed by observers worldwide and used in this research. We also thank the AAVSO HQ staff who have maintained the database over many decades, and the staff and others who have created and maintained the VSTAR time-series analysis package. This project was supported by the University of Toronto Work-Study Program, and the Dunlap Institute. We thank Professor Igor Soszynski and Matylda Soszynska for providing Figure 1.